# A Non-Invasive Load Monitoring Method for Edge Computing Based on MobileNetV3 and Dynamic Time Regulation

Hangxu Liu[1]
School of Information Science and Engineering
Fudan University
Shanghai, China
23210720103@m.fudan.edu.cn

Yaojie Sun*
School of Information Science and Engineering
Fudan University
Shanghai, China
yjsun@fudan.edu.cn

Yu Wang*
School of Information Science and Engineering
Fudan University
Shanghai, China
y_wangfdu@fudan.edu.cn

*Abstract*—In recent years, non-intrusive load monitoring (NILM) technology has attracted much attention in the related research field by virtue of its unique advantage of utilizing single meter data to achieve accurate decomposition of device-level energy consumption. Cutting-edge methods based on machine learning and deep learning have achieved remarkable results in load decomposition accuracy by fusing time-frequency domain features. However, these methods generally suffer from high computational costs and huge memory requirements, which become the main obstacles for their deployment on resource-constrained microcontroller units (MCUs). To address these challenges, this study proposes an innovative Dynamic Time Warping (DTW) algorithm in the time-frequency domain and systematically compares and analyzes the performance of six machine learning techniques in home electricity scenarios. Through complete experimental validation on edge MCUs, this scheme successfully achieves a recognition accuracy of 95%. Meanwhile, this study deeply optimizes the frequency domain feature extraction process, which effectively reduces the running time by 55.55% and the storage overhead by about 34.6%. The algorithm performance will be further optimized in future research work. Considering that the elimination of voltage transformer design can significantly reduce the cost, the subsequent research will focus on this direction, and is committed to providing more cost-effective solutions for the practical application of NILM, and providing a solid theoretical foundation and feasible technical paths for the design of efficient NILM systems in edge computing environments.

*Keywords*— Non-intrusive load monitoring, mobileNetV3, edge computing, dynamic time warping

## I. INTRODUCTION

Non-Invasive Load Monitoring (NILM) enables device-level power consumption analysis through integrated measurements at key metering points, such as home or industrial backbones. This approach eliminates the need for specialized single-device sensors while maintaining cost-effectiveness and deployment flexibility. Modern smart meters enable near real-time voltage/current measurements, enabling NILM-based power disaggregation without the need for a distributed metering infrastructure.

While traditional NILM applications focus on statistical data collection, emerging edge computing enables new types of services such as anomaly detection and security enhancements. State-of-the-art approaches utilize high-dimensional feature spaces and computationally intensive machine learning algorithms [1] to achieve significant disaggregation accuracies at the cost of large memory and processing requirements [2]. As shown in Fig. 1, traditional cloud architectures delegate computationally intensive feature extraction and classification tasks to back-end servers, which require high-bandwidth uplinks for meter-to-cloud data transfer.

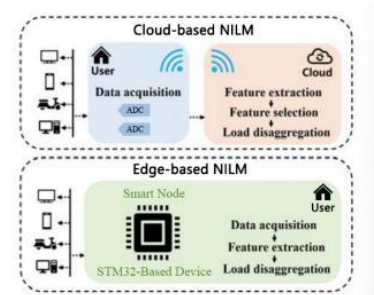

Figure 1: Non-Intrusive Load Monitoring Architecture

1) Cloud-Based 2) Edge-Based.

The edge computing framework effectively solves the scalability issues such as communication delay, bandwidth limitation and privacy protection faced by cloud computing in NILM applications [3]. Compared to the risk of user energy consumption data leakage that may result from centralized processing in the cloud [4], the edge computing scheme has the following advantages: 1) it provides real-time energy consumption feedback and supports instant energy-saving regulation and fault early warning; 2) empirical evidence shows that its energy-saving effect (5-20%) significantly outperforms that of the cloud-based scheme (0-10%) [5]. These advantages establish the technical value of edge computing in NILM applications. However, MCU-based edge deployments face serious challenges, as on-chip storage resources (SRAM/FLASH) are six orders of magnitude lower than in the cloud, making traditional algorithms difficult to port directly. There is an urgent need to develop memory-efficient lightweight computational frameworks that balance memory footprint, computational latency, and monitoring accuracy through optimized feature space design to achieve efficient NILM applications in resource-constrained environments. Meanwhile, edge NILM faces the contradiction between cutting-edge algorithms with high sampling frequency and memory requirements, and the lack of performance of commercial smart meter hardware. Although the existing research programs can meet the algorithm requirements, the high-end platforms are not suitable for low-cost edge deployment. To this end, this study develops customized MCU-based smart measurement nodes to optimize the simulation back-end for richer features. At the same time, we propose an innovative edge deployment method, which balances memory, latency and accuracy by adjusting feature dimensions to realize lightweight edge deployment of NILM algorithm. This paper promotes the transition of NILM from a high-end platform to an economical edge solution, with specific contributions as follows:

1) An optimized NILM framework for edge devices is proposed to achieve efficient load decomposition through key time-frequency domain feature extraction and classification modules;

2) Innovatively developed a time-frequency domain dynamic time regularization algorithm based on threshold current, which basically requires only current signal to complete load decomposition, effectively circumvents the MCU voltage acquisition accuracy limitation and reduces hardware cost;

3) The memory and performance requirements of the feature extraction stage are systematically evaluated, and the resource constraints (SRAM/Flash occupancy and execution cycle) of the worst-case model operation are determined;

4) By comparing multiple supervised learning algorithms, an edge NILM model based on MobileNetV3 is constructed, which significantly reduces computational complexity while maintaining 95% accuracy.

The structure of this paper is organized as follows: the second part systematically reviews the research progress of existing load decomposition methods; the third part elaborates the dynamic time regulation algorithm based on threshold current and the key technical aspects of the NILM framework; the fourth part demonstrates the experimental results and analyses of the feature extraction and load decomposition algorithms; and finally, it summarizes the research results and looks forward to the future research direction.

## II. Related Works

Since Hart et al [6] proposed the concept of non-intrusive load monitoring (NILM), early model-driven methods based on steady-state feature space performed well in simple switched load identification, but were not sufficiently capable of recognizing complex loads such as finite state machines and continuously variable devices. With the development of machine learning, J. Kelly et al [7] introduced neural networks into the field of NILM; convolutional neural networks have also been widely used, e.g., Hui Yin et al [8] utilized deep convolutional neural network fusion of spatio-temporal features to improve the accuracy of appliance identification, and Yanchi Liu et al [9] realized load classification based on voltage-current trajectory migration learning. However, these advanced methods are difficult to realize real-time applications on low-cost microcontroller units due to high computational complexity and memory requirements.

In recent years, research on edge-based non-intrusive load monitoring (NILM) has continued to advance, and several teams have achieved results: R. Gopinath et al [11] proposed DeepEdge-NILM for commercial buildings based on ESP32, borrowing feature fusion and LSTM to optimize similar load monitoring; Barsocchi et al [10] designed a low-cost smart meter based on an LED probe with the Arduino platform FSM to design low-cost smart meters for low-frequency feature monitoring of home power consumption; S. Kotsilitis et al [12] proposed a lightweight field event detection algorithm to achieve efficient steady-state identification with simple features and multiple criteria. However, existing schemes generally suffer from the lack of high-frequency analog front-ends and limited on-board resources, making it difficult to support advanced sensor processing, and often rely on third-party cloud services to obtain device-level information, which results in challenges such as data transmission costs, privacy risks, and increased system complexity, and greatly restricts the development of edge NILM technology.

## III. Proposed Method

The experimental program of this paper is explained in this section. First, an experimental platform needs to be built for data acquisition, which includes voltage and current data from five electrical appliances. Second, a combination of time and frequency domains is used and a dynamic time regularization algorithm is employed to extract detailed features of different cycles of current as the output of the neural network. Finally, the data is trained on five deep learning models and deployed on STM32H745 to realize non-intrusive load monitoring at the edge.

### A. A. Intelligent Measurement Node Designed with Dual MCU Architecture

In this study, a dual-MCU synergistic architecture is adopted for raw data acquisition and NILM algorithm deployment. The STM32G474 (Arm Cortex - M4 core, 170MHz, integrated 512KB Flash, data gas pedal and HR timer) is used as the front-end, which completes the data measurement by virtue of its high integration advantage, and transmits the results in real time to the STM32H745 through SPI communication. The latter is based on the high-performance Arm Cortex-M7/M4 dual-core architecture, equipped with FPU, complete DSP instruction set and MPU, integrating 2MB dual-area Flash and 1MB RAM, and combining with ADC, DMA and other peripherals, it undertakes the task of accelerating and parallelizing complex algorithms, and ultimately realizes the transmission of the data results to the terminal through the serial port.

To obtain raw voltage and current data, the Smart Measurement Node utilizes a Hall sensor, LT 58 - S7, and a voltage transformer, ZMPT101B. The LT 58 - S7 is a LEM current sensor with a range of 50A, a 50A rated primary RMS current, and a measurement range of 0 to $\pm 70A$. The ZMPT101B is a unidirectional AC output voltage mutual sensor module with inputs to measure AC voltages up to 250V and adjust the outputs as needed. The ZMPT101B is an input unidirectional AC output voltage mutual sensing module that can measure AC voltages up to 250V and adjust the output as required. The Hall sensor is connected to ADC1 of STM32G474, and the voltage transformer is connected to ADC2. In STM32CubeIDE, G474 and H745 are set up to transmit data through SPI communication, and at the same time, communicate with the computer through the USART interface to realize the real-time streaming transmission of data, which is convenient for the storage and analysis at the receiving end.

### B. Sampling frequency setting

After the hardware deployment is completed, the sampling frequency of the MCU's ADC is optimized based on the Nyquist Sampling Theorem (the sampling frequency needs to be at least two times the highest frequency of the signal to ensure a distortion-free signal restoration). Under the STM32CubeIDE environment, by adjusting the external clock configuration and the frequency divider and multiplier factors, the internal clock frequency of the ADC is set to about 21.714286 MHz, and combined with the sampling period setting of 810.5 clock cycles, and based on the formula for calculating the ADC sampling frequency (a function of the system-allocated clock frequency and the internal divider, and the sampling period), a final sampling frequency of about 6629.3Hz sampling frequency, to meet the experimental demand for a complete and distortion-free sampling signal.

$$f_{ADC\_sample} = \frac{f_{ADC\_dock}}{ADC\_Prescaler \times Sampling\_cycles} \quad (1)$$

The top of the formula indicates the clock assigned to the peripheral ADC in STM32CubeIDE, and the bottom of the formula is the ADC's dividing factor and sampling period, respectively.

### C. Hybrid DTW algorithm in time and frequency domains

In order to solve the load decomposition problem caused by the overlapping operation of multiple electrical appliances, this study proposes a non-intrusive load monitoring (NILM) method based on power threshold triggering and feature fusion. Specifically, the instantaneous power is calculated from the collected voltage and current data to construct the time domain feature, and then the current signal is converted to the frequency domain using the fast Fourier transform (FFT), which is then fused with the power feature to form a composite feature F. When the power value exceeds the preset threshold of 5W, the system triggers the event monitoring mechanism to extract the current sequence at specific times before and after the event (j, j-10, j-20 before the event and j+1, j+10, j+20 after the event). When processing these sequences, the dynamic time warping (DTW) algorithm based on the dynamic programming principle is used. This algorithm breaks through the limitation of the traditional Euclidean distance on the equal length of the sequence, and constructs the optimal matching path between the current sequences by dynamically adjusting the time axis, effectively dealing with problems such as phase offset and timing fluctuation, and significantly enhancing the robustness of load feature recognition. Compared with Euclidean distance, DTW can flexibly align sequences of unequal lengths, quickly calculate the optimal match and minimum distance, better meet the needs of NILM scenarios, and help achieve accurate load decomposition under complex working conditions. Usually, two current sequences of different lengths X=(x1,x2,...,xi) and Y=(y1,y2,...,yj) in a single cycle are obtained through time domain and frequency domain features, and their DTW distance can be calculated by the following recursive formula：

$$D(i,j) = d(i,j) + min\begin{cases} D(i-1,j) \\ D(i,j-1) \\ D(i-1,j-1) \end{cases} \quad (2)$$

d(i,j) represents the distance between each point of the two sequences. D(i-1,j) means that the i-th point of sequence X repeatedly matches the j-th point of Y (time axis stretching). D(i,j-1) means that the j-th point of sequence Y repeatedly matches the i-th point of X (time axis compression). D(i-1,j-1) means that Xi directly matches Yj (one-to-one alignment).

### D. Lightweight convolutional neural network model architecture

To meet the resource - constrained requirements of MCU - based deployment, this study leverages MobileNetV3 [20], a lightweight CNN designed for mobile and embedded devices. Compared with traditional networks like VGG16 and ResNet50 [13], MobileNetV3 enhances computational efficiency and inference speed via neural architecture search (NAS) and convolutional structure optimization. Its key components include hard activation functions (h - swish and h - sigmoid), which offer ReLU - like nonlinearity at low cost; Squeeze and Excitation (SE) modules that adaptively emphasize key features by compressing channel features and generating weighting coefficients; and depthwise separable convolutions that split standard convolutions to reduce parameters and computations. By integrating these elements, MobileNetV3[18] balances efficiency and performance while maintaining a lightweight design suitable for MCU deployment.

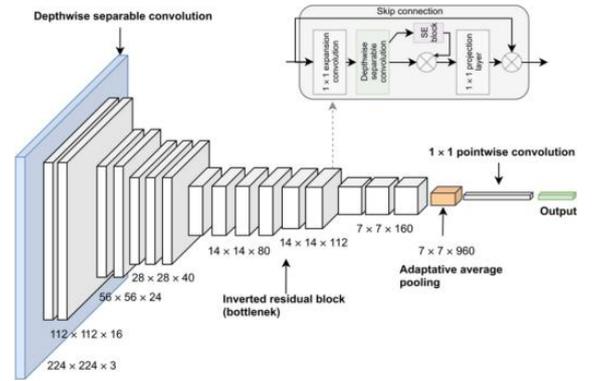

Figure2.MobileNetV3 structure diagram

Figure 2 shows the network architecture of MobileNetV3. This study uses its small version for experiments. At the network output layer, the model output is converted into probability distribution of each category through the softmax activation function, and the error is calculated with the true label through the multi-classification cross entropy loss function. With the help of the back-propagation algorithm, the network parameters are iteratively updated according to the loss value, and the model prediction accuracy is gradually optimized, so that its output is closer to the true label.

## IV. EXPERIMENT AND VALIDATION

This section tests and verifies the proposed non-intrusive load detection algorithm based on a lightweight neural network architecture through experiments. In order to comprehensively evaluate the performance of the algorithm, five typical power loads, namely laptop computers, refrigerators, washing machines, hair dryers, and incandescent lamps, are selected to conduct two case studies, focusing on verifying the effectiveness and accuracy of the algorithm deployed on the embedded terminal side. Before the experiment, the frequency domain features of the feature extraction module are optimized, and then case analysis is carried out: Case 1 conducts an independent test on a single load, and Case 2 runs through a combination of multiple loads to verify the comprehensive detection capabilities of the algorithm and model.

*A. Results of lightweight model optimization strategy*

This section discusses the results of the algorithm and lightweight model optimization strategy. First, the computational resources and memory usage of extracting time-frequency domain features on the ARM Cortex-M7 core of the STM32H745 are quantitatively analyzed. Then, through comparative experiments of two load monitoring scenarios, the runtime characteristics of different frameworks are evaluated to screen the optimal algorithm for edge non-intrusive load monitoring (NILM). Table 1 gives the memory consumption and execution cycle data of feature extraction of each component of the ARM Cortex-M7 core in detail. At the same time, combined with the memory constraints of the MCU device, the analysis results of the runtime SRAM memory and FLASH flash storage requirements are given.

In the feature extraction stage, the original ADC sampling values are first subjected to gain/offset calibration (RawConv). This program is used as a means of data calibration to calculate the calibrated current and voltage measurement values with the help of multiplicative gain and offset parameters, providing a basis for subsequent algorithms. The sample conversion of dual-channel current and voltage signals requires 16K cycles, while the single-channel current signal processing only requires 6.5K cycles. After obtaining the electrical parameters, the power characteristics are calculated on a 100ms time frame (more than 1000 sampling points). The extraction of active power P and apparent power S takes 18K and 12K cycles, respectively. The reactive power is calculated based on the obtained P and S values. The calculation overhead is affected by the working conditions. The best and worst cases are 72 cycles and 40K cycles, respectively.

For the 50Hz harmonic frequency in the home scenario, this study calls the ARM CMSIS-DSP software library FFT routine in the STM32CubeIDE environment to extract 50Hz odd current harmonics, and monitors the SRAM and FLASH storage requirements in real time during operation. FFT spectrum processing generates 120 real and imaginary components in a single frame, and the calculation takes about 90K cycles, which requires 60KB Flash to store the rotation coefficients and bit reversal lookup table; if the FFT component reordering is skipped, the calculation time can be optimized to 84K cycles. According to calculations, the total computational overhead of the feature extraction link is about 140K cycles under the worst working condition, and the storage requirements are 60KB FLASH and 125KB SRAM. In the experiment, the main frequency of the STM32H7 MCU is configured to 72MHz, and the 100ms time window corresponds to 6.5Mcycles. Feature extraction only accounts for 1.94% of the total cycle (140K cycles), and the remaining 71.86Mcycles provide sufficient computing resources for the deployment of the load identification model.

TABLE I

| Load combination | SRAM(kB) | Flash(kB) | Cycles(K) |
|---|---|---|---|
| RawConv(V＆I） | 6 | 4 | 16.5 |
| RawConv(I） | 4 | 4 | 6.5 |
| P | 4 | 12 | 18 |
| S | 5 | 12.5 | 12 |
| FFT | 10 | 15.6 | 90 |
| FFT(Skip reordering) | 8 | 10.2 | 40 |

Quantitative analysis based on table data shows that the optimized FFT algorithm has achieved significant breakthroughs in storage and computing efficiency: SRAM usage is reduced by 2KB, Flash storage requirements are reduced by about 34.6%, and the running time is shortened by 55.55% compared to before optimization, effectively alleviating the resource constraints of the MCU platform. In addition, this study applied the optimization strategy to load model verification, and the experimental results further confirmed its effectiveness and engineering practicality.

*B. Case1：Single Load Identification Test*

Case 1 focuses on the scenario where only a single load is operated within a single period of time. The study uses ADC to collect current and voltage data of common household appliances such as laptops, refrigerators, washing machines, hair dryers, and incandescent lamps. 10,000 sets of raw data are selected for each appliance. By manually controlling the switches of the electrical appliances, the system collects waveform data when a single load is running and an event occurs, and builds a complete data feature library. After the raw data is processed by the feature extraction algorithm, the DTW distance of similar currents is used as the input of the neural network, and the training set, validation set, and test set are divided into 70%, 10%, and 20% ratios. After training and verification of the neural network model, the characteristic curves of 5 types of load switching behaviors are obtained, and then the load monitoring analysis of the lightweight convolutional neural network is carried out based on the two-dimensional features.

After completing the model training and testing, it is converted to ONNX format and deployed to STM32H745 with the help of CubeAI in STM32CubeIDE. Experiments were conducted on various loads, and Figure 3 shows the specific test results. Taking a laptop as an example, the algorithm and model recognition method proposed in this study determines that the probability that the test data belongs to a laptop is 99.5%, and there is only a 0.5% probability of misjudging it as an incandescent lamp, clarifying the load type. In addition, as can be seen from Figure 3, the recognition accuracy of various types of loads exceeds 97.5%, which fully verifies the effectiveness of the proposed algorithm and lightweight model.

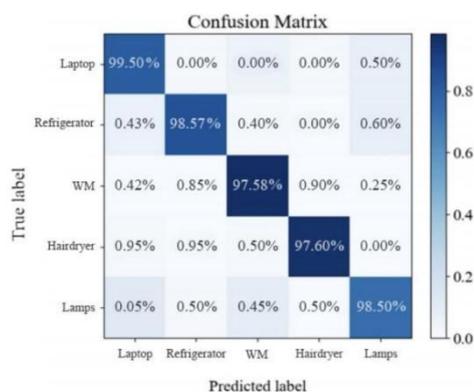

Figure3.Identification accuracy result of one experiment.

## C. Case2：Multi-Load Identification Test

Case 2 focuses on the complex working condition of multiple loads running in parallel, aiming to verify the adaptability of the proposed method in multi-load scenarios. Given that there are as many as 31 different load combinations, due to space limitations, Table 2 selects several typical load combination examples to illustrate the effectiveness of the algorithm and model.

TABLE II
THE ACCURACY OF MULTI-LOAD IDENTIFICATION TEST BEFORE OPTIMIZATION

| Load combination | Accuracy | Precision |
|---|---|---|
| Laptop＋Refrigerator | 92% | 95% |
| Washing machine＋Laptop | 95% | 98% |
| Hairdryer＋Washing machine | 95% | 97% |
| Hairdryer＋Laptop | 96% | 99% |
| Lamps＋Hairdryer | 95% | 98% |

THE ACCURACY OF MULTI-LOAD IDENTIFICATION TEST USING OPTIMIZATION

| Load combination | Accuracy | Precision |
|---|---|---|
| Laptop＋Refrigerator | 91.5% | 94.5% |
| Washing machine＋Laptop | 95.5% | 97% |
| Hairdryer＋Washing machine | 94% | 98% |
| Hairdryer＋Laptop | 96% | 99% |
| Lamps＋Hairdryer | 95% | 98% |

The data in the table presents the accuracy and precision indicators of the multi-load identification test under different load combinations before and after optimization. The test results show that the overall performance of various indicators is excellent, and the accuracy and precision are mostly above 95%, which fully verifies the effectiveness and reliability of the algorithm model in this study.

## D. Case3：Comparison With Other Methods

This study conducted comparative experiments on the proposed load identification algorithm based on lightweight convolutional neural network and common machine learning and deep learning models such as random forest [14], support vector machine [15], LSTM [16], MobileNetV1 [17], and MobileNetV2 [19]. Multiple rounds of tests were conducted on the same load, and a systematic analysis was carried out from the perspectives of accuracy, precision, and F1 score. The specific results are shown in Table 3. The experimental results fully verified the effectiveness and superiority of the proposed algorithm.

TABLE III
COMPARISON RESULTS AMONG OTHER DEEP LEARNING METHODS

| Models | Accuracy | Precision | F1-score |
|---|---|---|---|
| RF | 83.00% | 87.25% | 85.50% |
| SVM | 85.00% | 88.50% | 86.25% |
| LSTM | 91.00% | **95.75%** | 92.25% |
| MobileNetv1 | 90.5% | 91.75% | 91.50% |
| MobileNetv2 | 93.00% | 93.95% | 92.75% |
| **MobileNetv3** | **95.00%** | 95.55% | **94.50%** |

As can be seen from Table 3, the accuracy of the machine learning and deep learning methods involved in the comparison is over 83%, while the load recognition algorithm proposed in this paper has an accuracy of 95%, which is better than the other models. In terms of recognition accuracy, this paper's algorithm ranks second with 95.55%, which is only 0.2% lower than the best LSTM model, and has the best F1-score performance. Combining the results of all comparisons, the proposed method in this paper has better recognition performance for most residential load data.

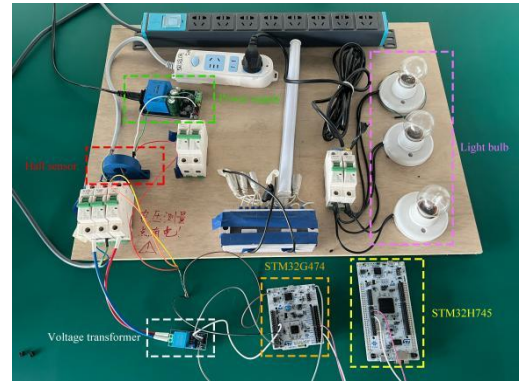

Figure4. A NILM monitoring device in lab.

## E. Discussion About the Practical Application

In order to realize the engineering application of the proposed method, as shown in Fig. 4, a non-intrusive load monitoring (NILM) device is constructed in this study. The device is based on the STM32G474 and STM32H745 MCU motherboards as the core, and integrates Hall current sensors and voltage transformers, which perform the functions of data acquisition and processing, current detection and voltage measurement, respectively. At present, the NILM device and supporting algorithms have completed the initial deployment and testing in the laboratory. Given that the project is in the early stage of research and development, we will focus on optimizing the recognition function of household appliances and promoting the transformation of the technology into practical application scenarios.

## V. CONCLUSIONS

This study proposes a non-intrusive load monitoring (NILM) framework for edge computing that incorporates optimized MobileNetV3 architecture and time-frequency hybrid dynamic time regularization (DTW) algorithms, which effectively solves the deployment challenges of traditional NILM methods in resource-constrained environments of microcontroller units (MCUs). Experimentally verified, the framework achieves up to 95% load identification accuracy on STM32 microcontrollers, outperforming traditional machine learning methods in both single-load and multi-load scenarios, providing a feasible solution for low-cost edge energy monitoring. Meanwhile, the frequency domain feature extraction optimization reduces the execution time by 55.55% and memory consumption by about 34.6%, which significantly improves the operation efficiency of the algorithm under resource-constrained conditions. The basic current-based DTW algorithm successfully breaks through the bottleneck of load identification and decomposition when multiple appliances overlap by fusing the time-frequency domain features, and enhances the robustness of load feature identification in complex scenarios. Future research will focus on neural architecture search and adaptive sampling optimization to further enhance the monitoring efficiency and accuracy of the NILM framework in edge computing environments.

## Acknowledgments

This work is supported by the Science and Technology Project of State Grid Shanghai Electric Power Company (No. B3094024000L). We deeply appreciate the financial and resource - related support provided by the company, which has played a crucial role in promoting the progress of this research.

## REFERENCES

[1] T. Bernard, M. Verbunt, G. v. Bögel and T. Wellmann, "Non-Intrusive Load Monitoring (NILM): Unsupervised Machine Learning and Feature Fusion: Energy Management for Private and Industrial Applications," International Conference on Smart Grid and Clean Energy Technologies (ICSGCE 2018): Kajang, Malaysia, Sg. Long, Cheras, Kajang, Malaysia, 29 May-1 June, 2018, pp. 174–180, doi: 10.1109/ICSGCE.2018.8556735.

[2] K. -K. Kee, Y. S. Lim, J. Wong and K. H. Chua, "Non-Intrusive Load Monitoring (NILM) – A Recent Review with Cloud Computing," 2019 IEEE International Conference on Smart Instrumentation, Measurement and Application (ICSIMA), Kuala Lumpur, Malaysia, 2019, pp. 1-6, doi: 10.1109/ICSIMA47653.2019.9057316.

[3] Christos Stergiou, Kostas E. Psannis, Brij B. Gupta, Yutaka Ishibashi,Security, privacy & efficiency of ustainable Cloud Computing for Big Data & IoT,Sustainable Computing: Informatics and Systems, olume 19,2018,Pages 174-184,ISSN 2210-5379,https://doi.org/10.1016/j.suscom.2018.06.003.

[4] H. Cao, S. Liu, L. Wu, Z. Guan, and X. Du, "Achieving differential privacy against non-intrusive load monitoring in smart grid: A fog computing approach," in Concurrency and Computation: Practice and Experience, Wiley, vol. 31, issue 22, no. e4528, 2019, doi: 10.1002/cpe.4528.

[5] Sebastien Houde, Annika Todd, Anant Sudarshan, June A. Flora , and K. Carrie Armel, "Real-time Feedback and Electricity Consumption: A Field Experiment Assessing the Potential for Savings and Persistence," The Energy Journal, International Association for Energy Economics, vol. 34, no. 1, 2013, pp. 1944-9089, doi:10.5547/01956574.34.1.4.

[6] G. W. Hart, "Nonintrusive appliance load monitoring," in Proceedings of the IEEE, vol. 80, no. 12, Dec. 1992, pp.1870-1891, doi: 10.1109/5.192069.

[7] J. Kelly and W. Knottenbelt, "Neural NILM: Deep neural networks applied to energy disaggregation", ACM BuildSys, Seoul, South Korea, 2015, pp. 55-64, doi: 10.1145/2821650.2821672.

[8] H. Yin, K. Zhou and S. Yang, "Non-Intrusive Load Monitoring by Load Trajectory and Multi-Feature Based on DCNN," in IEEE Transactions on Industrial Informatics, vol. 19, no. 10, pp. 10388-10400, Oct. 2023, doi: 10.1109/TII.2023.3240924.

[9] Y. Liu, X. Wang and W. You, "Non-Intrusive Load Monitoring by Voltage–Current Trajectory Enabled Transfer Learning," in IEEE Transactions on Smart Grid, vol. 10, no. 5, pp. 5609-5619, Sept. 2019, doi: 10.1109/TSG.2018.2888581.

[10] P. Barsocchi, E. Ferro, F. Palumbo and F. Potortì, "Smart meter led probe for real-time appliance load monitoring,"SENSORS, IEEE, Valencia, 2014, pp. 1451-1454, doi: 10.1109/ICSENS.2014.6985287.

[11] R. Gopinath, Mukesh Kumar,DeepEdge-NILM: A case study of non-intrusive load monitoring edge device in commercial building,Energy and Buildings,Volume 294,2023,113226,ISSN 0378-7788, https://doi.org/10.1016/j.enbuild.2023.113226.

[12] S. Kotsilitis, E. Kalligeros, E. C. Marcoulaki and I. G. Karybali, "An Efficient Lightweight Event Detection Algorithm for On-Site Non-Intrusive Load Monitoring," in IEEE Transactions on Instrumentation and Measurement, vol. 72, pp. 1-13, 2023, Art no. 9000313, doi: 10.1109/TIM.2022.3232169.

[13] S. Mascarenhas and M. Agarwal, "A comparison between VGG16, VGG19 and ResNet50 architecture frameworks for Image Classification," 2021 International Conference on Disruptive Technologies for Multi-Disciplinary Research and Applications (CENTCON), Bengaluru, India, 2021, pp. 96-99, doi: 10.1109/CENTCON52345.2021.9687944.

[14] Rigatti, S. J. (2017). Random forest. Journal of Insurance Medicine, 47(1), 31-39.

[15] Xue, H., Yang, Q., & Chen, S. (2009). SVM: Support vector machines. In The top ten algorithms in data mining (pp. 51-74). Chapman and Hall/CRC.

[16] Yu, Y., Si, X., Hu, C., & Zhang, J. (2019). A review of recurrent neural networks: LSTM cells and network architectures. Neural computation, 31(7), 1235-1270.

[17] Mijwil, M. M., Doshi, R., Hiran, K. K., Unogwu, O. J., & Bala, I. (2023). MobileNetV1-based deep learning model for accurate brain tumor classification. Mesopotamian Journal of Computer Science, 2023, 29-38.

[18] A. Howard et al., "Searching for MobileNetV3," 2019 IEEE/CVF International Conference on Computer Vision (ICCV), Seoul, Korea (South), 2019, pp. 1314-1324, doi: 10.1109/ICCV.2019.00140.

[19] K. Dong, C. Zhou, Y. Ruan and Y. Li, "MobileNetV2 Model for Image Classification," 2020 2nd International Conference on Information Technology and Computer Application (ITCA), Guangzhou, China, 2020, pp. 476-480, doi: 10.1109/ITCA52113.2020.00106.